\documentstyle[prl,aps]{revtex}
\twocolumn
\begin{document}
\input{epsf}
\draft
\twocolumn[\hsize\textwidth\columnwidth\hsize\csname@twocolumnfalse\endcsname
\title{Logarithmically Slow Expansion of Hot Bubbles in Gases}
\author{Baruch Meerson and Pavel V. Sasorov \cite{adr}}
\address{Racah Institute of Physics, Hebrew
University   of  Jerusalem,
Jerusalem 91904, Israel}
%\author{Pavel V. Sasorov}
%\address{Institute of Theoretical and Experimental Physics, Moscow Russia}
\author{Ken Sekimoto}
\address{Yukawa Institute for Theoretical 
Physics,  Kyoto University, Kyoto 606-8502, Japan}
\maketitle
\begin{abstract}
We report logarithmically slow expansion of hot bubbles
in gases in the process of cooling.  
A model problem first solved, when the 
temperature has compact support. Then
temperature
profile decaying exponentially at large distances is considered. 
The periphery of the bubble is shown to
remain essentially 
static (``glassy") in the process of cooling until it is taken over by a
{\it logarithmically\,}  slowly expanding ``core". An analytical
solution to the problem  is obtained by matched asymptotic
expansion. This problem gives an example of how logarithmic
corrections enter dynamic scaling.
\end{abstract}
\vskip1pc]
\narrowtext
%\pagebreak
Dynamic scaling behavior of extended nonlinear
systems out
of equilibrium have attracted much attention in different
areas of physics \cite{scaling}. In  
continuum models dynamic scaling is intimately related to 
self-similar asymptotics of nonlinear partial 
differential equations \cite{Barenblatt}. Sometimes
logarithmic corrections
enter dynamic scaling laws \cite{scaling}. No general scenario
for their appearance is known.  One can say from experience that
they appear in marginal cases, dividing (in an appropriate
functional space) regimes with qualitatively different
behavior. The aim
of this work is to investigate one particular setting, of
general interest,
where logarithmic corrections to scaling appear somewhat unexpectedly: 
cooling dynamics of hot bubbles in gases.

Heat transfer in gases, 
strongly heated locally, looks quite differently
from the simple picture provided by the linear
heat diffusion equation.
The difference is mainly due to the small-Mach-number 
gas flow that develops
(even at zero gravity) owing to intrinsic pressure gradients.
This 
conductive cooling flow
(CCF) brings in cold
gas from the periphery and strongly modifies the cooling dynamics. Some 
aspects of CCFs
have been studied 
experimentally and
theoretically, mainly in the context of 
the late stage of strong explosions \cite{M,K,GLM}. In this 
Letter we report a new, striking feature of a CCF.  We find that, if 
the initial temperature profile
decays rapidly enough at large distances 
[like $\exp (-k|x|)\,, k>0$],
the hot bubble, while cooling down significantly, 
expands {\it logarithmically} slowly. 

Starting from the continuity, momentum and
energy equations for an inviscous ideal gas at zero gravity, 
and employing the small-Mach-number expansion,
one arrives at the following nonlinear equation for the scaled
gas temperature \cite{M}:
\begin{equation}
\partial_t T = T^2 \partial_x(T^{\nu-1} \partial_x T)\,,
\label{1}
\end{equation}
where indices $t$ and $x$ stand for partial derivatives (a slab geometry
is assumed), and $\nu$ is the
exponent in the power-law temperature dependence of 
the heat conductivity of the gas \cite{density}.

The scaled 
gas pressure stays constant (and equal to unity) in this approximation, so
the scaled gas density 
is simply $\rho(x,t) = T^{-1}(x,t)$, while the gas velocity is 
$v(x,t) = T^{\nu} \partial_x T$ \cite{M}.
Therefore, once solving Eq. (\ref{1}) for 
the temperature, one can easily find all other variables. 

Eq. (\ref{1})
has a multitude of similarity solutions:
\begin{equation}
T_{\beta}(x,t) = t^{\frac{2\beta-1}{\nu+1}} \theta(x/t^{\beta}) \,,
\label{2}
\end{equation}
where $\beta$ is an arbitrary 
parameter. Therefore, an interesting selection problem
appears, like in many other situations in nonlinear dynamics of
extended systems \cite{Barenblatt,CH}.

Eq. (\ref{1}) has appeared 
in the context of cooling of the ``fireball"
produced by a
strong local explosion in a gas \cite{M,K,GLM}. An explosion 
involves energy release 
on a time scale short compared to the characteristic acoustic
time. In this case the 
preceding rapid stage
of the dynamics produces an
inverse power-law dependence of the gas temperature
on the distance from the explosion site \cite{ZR}. It has been shown
\cite{M,K} that the exponent
of this  power law uniquely
selects the scaling 
exponent $\beta$. As the result, the 
fireball expansion exhibits a power law in time. 

A different type of local gas heating occurs when the 
time scale
of the energy release is long compared to the acoustic time, but still
short compared to the cooling time. In this case the initial
temperature
profile is more localized, as it reflects the
spatial structure of the heating agent (for example, the radial intensity of 
laser beam). This 
regime will be in the focus
of this Letter. We will see that there is
{\it no} self-similar asymptotics to this problem. Instead, the solution
approaches, at long times, a ``quasi-similarity" asymptotics with logarithmic
corrections to scaling. 

Consider first a model problem when the
initial temperature profile 
has compact support: $T(x,0)=T_0(x)>0$ at 
$x \in [-L,L]$, and zero elsewhere. We 
will limit ourselves
to a temperature-independent heat conductivity, $\nu=0$. 
Despite this choice, the nonlinearity 
of Eq. (\ref{1}) persists. Assume
symmetry with respect to $x=0$  and impose the Neumann
boundary conditions: $\partial_x T(0,t)= \partial_x T(L,t)=0$.
A local analysis
of Eq. (\ref{1}) near the edge of support of its solution $T(x,t)$
shows that 
the support remains compact and {\it unchanged} for $t>0$. Therefore, 
this model problem exhibits a {\it complete} localization. What is the 
late-time behavior of the temperature?
The constancy of support immediately selects $\beta=0$,
so the similarity ansatz becomes $T_0(x,t)= t^{-1} \theta (x)$.
Then Eq. (\ref{1}) yields 
$\theta (x) = (a^2/2) \cos^2 (x/a)$ for $x \in [-L,L],  
\theta (x) = 0$ elsewhere, and $a=\pi L/2$. This simple 
similarity solution describes cooling of
the hot bubble
(and filling it with the dense gas) without
{\it any} change in the bubble size.

Remarkably, $T_0(x,t)$ represents a long-time
asymptotics for {\it any} initial condition that 
has compact support $[-L,L]$ and obeys the Neumann boundary 
conditions. We will show here only that
this solution is linearly 
stable with respect to small perturbations, and
find the spectrum of the linearized problem.
Introduce new variables: $u=t\,T(x,t)$  and  $\tau=\ln t$.
Eq. (\ref{1}) assumes the form
\begin{equation}
\partial_{\tau} u = u - (\partial_x u)^2 + u \partial_{xx} u\,,
\label{3}
\end{equation}
while the similarity solution $T_0 (x,t)$ becomes
steady-state solution $\theta (x)$. Introducing
a small correction $v(x,\tau)$ to this solution 
and linearizing Eq. (\ref{3}), we 
obtain $\partial_{\tau} v = \hat{L} (\xi) v$, where
\begin{equation}
\hat{L}(\xi) = (1/2) \cos^2 \xi\, \partial_{\xi\xi} + 
\sin 2\xi \,\partial_{\xi} +
2 \sin^2 \xi
\label{4}
\end{equation}
and $\xi=x/a$. We look for the eigenfunctions in the form of
$v(\xi,\tau) = e^{ \gamma \tau} \psi_\gamma (\xi)$. The general solution of
the resulting ordinary differential equation is
\begin{eqnarray}
\psi_\gamma (\xi) = C_1 \cos^2 \xi\,  _2F_1 (a_-, a_+, 1/2, -\tan^2 \xi) 
\nonumber\\
+\, C_2 \cos \xi \sin \xi\, _2F_1 (b_-, b_+, 3/2, -\tan^2 \xi)\,,
\label{6}
\end{eqnarray}
where $_2F_1$ is the hypergeometric function, $C_1$ and $C_2$ are
arbitrary constants, $a_{\pm} = \left[1 \pm (8\gamma+9)^{1/2}\right]/4$ and
$b_{\pm} = \left[3 \pm (8\gamma+9)^{1/2}\right]/4$. Requiring 
that the perturbation remains small
compared to the unperturbed solution [and 
hence vanishes like $(\pi/2-\xi)^2$
or faster
at $\xi \to \pi/2$], we find the (continuous) spectrum
of the linearized problem: $-\infty <\gamma \le -1$. This result proves 
linear stability of the similarity solution $T_0 (x,t)$. Notice 
the presence of gap between the 
upper edge of the spectrum
$\gamma=-1$ and  stability border
$\gamma=0$. Going back to physical variables, we find that
small {\it temperature} perturbations around the similarity
solution exhibit a power law decay $t^{\gamma-1}$.

At this stage we notice that $\beta=0$ is a marginal
case dividing two qualitatively different types of dynamics as
described by the family of 
similarity solutions (\ref{2}). Indeed, solutions
with
$\beta>0$ describe 
power-law {\it expansions} \cite{M,K,GLM}, while
solutions with $\beta<0$ correspond to 
power-law {\it shrinkings} \cite{shrinking}. 
One can expect logarithmic corrections to 
appear ``on the background" of
the special case $\beta=0$, when the initial condition does {\it not}
have compact support, but decays rapidly at large distances. Therefore,
we assume that the initial temperature profile of the bubble is 
symmetric with respect to $x=0$ and
decays exponentially at
large distances \cite{intermediate}. We will continue using the
new variables and require  $u(x,0)\to c \exp (-k |x|)$
at $|x|\to\infty$, where $k$ and $c$ are positive constants. 
One can always put $k=1$ \cite{k=1}. We will be interested in a long-time
asymptotics of the solution: $\tau \gg 1$.
Our first important observation is that $u (x,\tau) = c \exp(\tau-x)$ is
an {\it exact} solution of Eq. (\ref{3}).  This traveling wave solution 
with a unit speed 
corresponds to a steady-state solution $T(x)=c \exp(-x)$
in physical variables, and it represents the correct asymptotics
of the solution to our problem at $x \to +\infty$.
What about the bubble 
``core"? We will show that it 
can be described, at $\tau \gg 1$, by a ``quasi-similarity"
solution plus small corrections: 
\begin{equation}
u (x,\tau) = 
u_0(x,\tau) + u_1(x,\tau) + \dots\,,
\label{expansion}
\end{equation}
where
\begin{equation}
u_0 (x,\tau) = \frac{a^2 (\tau)}{2}\, \cos^2\, \frac{x}{a(\tau)}
\label{QS2}
\end{equation}
and $ \dots \ll u_1 \ll u_0$. 
One of our goals is to find an
asymptotic expansion for $a(\tau)$. 

The leading term of $a(\tau)$
can be guessed immediately. Indeed,  
expansion of $u_0 (x,\tau)$ in 
powers of $x-\pi a(\tau)/2$ near the point $x=\pi a(\tau)/2$ begins with
the term 
$(1/2)\left(x-\pi a(\tau)/2\right)^2$. This is a wave traveling with speed
$\pi\dot{a}/2$ along the $x$-axis! Therefore, it is natural to look for
a {\it general} traveling wave solution
$v(x,\tau)=V(x-\tau)$ of Eq. (\ref{3}) with a unit speed and require that
it behaves like $(z+\mbox{const})^2/2$ at $z\to -\infty$ and
like $c\, \exp (-z)$ at $z\to +\infty$, where $z=x-\tau$. 
If such a solution  exists, we can match
it with the leading term of the quasi-similarity solution (\ref{QS2}) 
in the region 
$1\ll - (x-\pi a/2) \ll a, 1 \ll -z$, once
\begin{equation}
a(\tau) = 2 \tau/\pi\,.
\label{leading}
\end{equation}
Eqs. (\ref{QS2}) and 
(\ref{leading}) have important implications. First,
the temperature scaling with physical time $t$ at the 
bubble center acquires a logarithmic correction.
Second, the
bubble core expands logarithmically
slowly. We will 
show in the following that these are indeed correct results, calculate the
subleading and sub-subleading terms for $a(\tau)$, and 
find other attributes
of asymptotic solution. 

The general traveling wave solution of Eq. (\ref{3}),
$V(z)$, obeys the second-order equation
\begin{equation}
-V_z=V-V_z^2+VV_{zz}\,
\label{TR2}
\end{equation}
that is soluble analytically. One integration yields
\begin{equation}
V^{-1} (d V/d z)=-1-W\left(-\exp\left(-1-V^{-1}\right)\right)\,,
\label{TR8}
\end{equation}
where $W(\eta)$ is the product log function
defined as the
solution of equation $W e^W =\eta$ 
(see, {\it e.g.,} Ref. \onlinecite{Mathematica}, p. 751). The 
arbitrary constant in Eq. (\ref{TR8}) has been chosen to satisfy the 
required asymptotic behavior $V(z) \to (z+\mbox{const})^2/2$ at 
$z \to -\infty$. Notice that, as
$V>0$ and $V_z<0$, we should work with the negative
branch of the product log function: $\eta<0$ and $W(\eta) <0$.

Integrating Eq. (\ref{TR8}), we obtain the traveling wave
solution in an implicit form:
\begin{equation}
J(V)+ z + C = 0\,,
\label{TW}
\end{equation}
where
\begin{equation}
J(V) = \int\limits_1^{U(V)}\frac{d\zeta}{1-\zeta-e^{-\zeta}}\ ,
\label{TR12}
\end{equation}
$U(V)=-\ln\left[-W\left(-\exp (-1-V^{-1})\right)\right]$ 
and $C$ is an arbitrary constant.

To understand the asymptotic behavior of this
solution at $z\to -\infty$ and $z \to +\infty$, we 
need to know the asymptotics of $J(V)$.
After some algebra we obtain
\begin{equation}
J(V)= \left\{\begin{array}{lcl}
\ln V +\Delta_1 +{\cal O}\left(V
\right)\,,
\\
(2V)^{1/2}+\frac{1}{3} \ln V + \Delta_2 +
{\cal  O}\left(V^{-1/2}\right)\,,
\end{array}\right.
\label{TR28}
\end{equation}
at $V\to+0$ and $V\to +\infty$, respectively. Here
$$\Delta_1=\int\limits_1^\infty \frac{\left(1- e^{-\zeta}\right) 
d\zeta}
{\zeta (1-\zeta-e^{-\zeta})} = -1.46074400\dots\,,$$
$\Delta_2 = -2 -(1/3) \ln \,(2 e)- \Delta_3$, and
$$
\Delta_3 = \int\limits_0^1 
\left(\frac{1}{1-\zeta-e^{-\zeta}} +\frac{2}{\zeta^2}
+\frac{2}{3\zeta}\right) d\zeta = -0.05361892\dots
$$
Using Eqs. (\ref{TW}) and (\ref{TR28}) we obtain:
\begin{equation}
V(z)= e^{-z-C-\Delta_1} + {\cal O}\left(e^{-2z}\right)\quad\mbox{at}\quad
z\to+\infty
\label{TR30}
\end{equation}
and
\begin{eqnarray}
V(z)&=&(1/2) (z+C+\Delta_2)^2 \nonumber \\
& &+ \,(2/3) (z+C+\Delta_2)\, 
\ln \left(|z+C+\Delta_2|)/\sqrt{2}\right) \nonumber\\
& & +\, {\cal O}\left(\ln^2|z+C+\Delta_2|\right)
  \quad\mbox{at}\quad z\to-\infty\,.
\label{TR31}
\end{eqnarray}
The required asymptotic behavior
$V\to c\, \exp (-z)$ at $z\to +\infty$ selects $C= -\Delta_1 - \ln c$,
so the traveling wave solution (\ref{TW}) is now fully determined. 
After some rearrangement, we rewrite the asymptotics (\ref{TR30})
and (\ref{TR31}) as
\begin{equation}
V (z)=c e^{-z}+{\cal O}(e^{-2 z})
\mbox{~~~~~at~~~~~}
z\to+\infty\,,
\label{TR36}
\end{equation}
and
\begin{eqnarray}
V(z)&=&(1/2)(z-\Delta)^2+(2/3)\,
(z-\Delta)\, \ln |z-\Delta| \nonumber \\
& &+\, {\cal O}(\ln^2|z|)
\mbox{~~~~~at~~~~~} z\to-\infty\,,
\label{TR37}
\end{eqnarray}
where $\Delta=\Delta_1 - \Delta_2 +(1/3) \ln 2+\ln c$.

The leading term in Eq. (\ref{TR36}) corresponds to a steady-state 
solution in the physical variables, while
the subleading term  is
{\it exponentially} small with respect to the leading one. This
essentially static (``glassy") behavior of the solution at large
distances reflects {\it effective}
diffusion choking at small temperatures.

Now let us return to the bubble core description, Eq. (\ref{QS2}). Our 
basic assumption 
here
(supported by the results) is that, in the asymptotic stage $\tau\gg 1$, the
terms
$u_0, u_1, \dots$ depend on time
only through the time dependences of $a$, of $a$ and $\dot{a}$, of $a, \dot{a},
\dots$, respectively. The small parameter of 
this expansion is
$\dot{a}/a$.
In the zeroth approximation of this
perturbation scheme, $u_0$ obeys Eq. (\ref{3}) without
the time derivative term. In the first approximation  we obtain the following
linear equation:
\begin{equation}
\hat{L} \, u_1(\xi) = a \dot{a}\, \cos^2 \xi\, (1+ \xi \tan \xi)\,,
\label{QS6}
\end{equation}
where we have again used  $\xi = x/a$.

The zero modes of the operator $\hat{L}$ are
$\Upsilon (\xi) = \cos^2\xi + \xi\, \cos\xi\, \sin \xi$ and
$\Phi (\xi) = \sin\xi\, \cos\xi$. Looking for the general 
solution of Eq. (\ref{QS6}) in the form of 
$u_1=C_1(\xi)\, \Upsilon(\xi)+C_2(\xi)\, \Phi(\xi)$ and
defining $a(\tau)$ by the condition $u(0,\tau) =a^2 (\tau)/2$, we arrive
at
\begin{eqnarray}
u_1 (\xi) &=&-(2/3)\, a\dot{a}\, \cos^2 \xi\,
[(\xi\tan\xi-1)\, \ln\cos\xi \nonumber \\
& &+\, 2\ln (2/e)\, \xi\tan\xi +
\tan\xi\, \mbox{Im}\, \left[\mbox{Li$_2$}(-e^{2i \xi})\right]]\,,
\label{QS10}
\end{eqnarray}
where
$\mbox{Li$_2$}(x)=\sum_{k=1}^\infty k^{-2} x^k$ is the 
dilogarithm (see, {\it e.g.,} Ref. \onlinecite{Mathematica}, p. 743).
In the vicinity of $\xi=\pi/2$
\begin{equation}
u_1=\frac{\pi}{3}\, a\dot{a}\, \tilde{\xi}\,
\ln\frac{4\,|\tilde{\xi}|}{e^2} + a \dot{a}\,
{\cal O}\left(\,|\tilde{\xi}|^3
\ln|\tilde{\xi}|\right )\,,
\label{QS11}
\end{equation}
where $\tilde{\xi}=\xi-\pi/2$. Too close 
to $\xi=\pi/2$ the correction $u_1$ and its derivatives become
larger than the zero-order solution $u_0$ and its corresponding
derivatives, so the
perturbation procedure breaks down. Therefore, the bubble core solution
[Eqs.
(\ref{expansion}), (\ref{QS2}) and (\ref{QS10})]
should be matched with the traveling wave solution [Eq. (\ref{TW})]
in the region
where $|x-\pi a/2|$ is small enough (so that the leading
term of the asymptotics of $u_0$ is much larger than the subleading terms)
but, on the other hand, large enough (so that $u_1$ is small compared 
to $u_0$). Working in this region and
collecting the leading contributions from $u_0$ and $u_1$, we obtain 
after some
rearrangement:
\begin{eqnarray}
u&=&\frac{1}{2}\, \tilde{x}^2\,
+\frac{\pi}{3}\dot{a}\,\tilde{x} \,\ln|\tilde{x}| +
\frac{1}{a^2}\, {\cal O}\left(\tilde{x}^4\right) 
+\, \frac{\dot{a}}{a^2}\,{\cal O}\left(|\tilde{x}|^3\ln |\tilde{x}|\right)\nonumber \\
 & & + \,{\cal O}\left(\dot{a}^2\ln^2a\right)\, {\cal O}\left(\ln
|\tilde{x}|\right)
+\dots\,,
\label{QS14}
\end{eqnarray}
where
$\tilde{x} = x-(\pi/2) \, a - (\pi/3)\,\dot{a}\, \ln(e^2 a/4)$.

Now we can perform the matching procedure. We require that 
the $z\to -\infty$
asymptotics
of the traveling wave solution, Eq. (\ref{TR37}), coincide
with the asymptotics (\ref{QS14})
of the bubble core solution. This yields
\begin{equation}
a(\tau)=\frac{2}{\pi k}\, \left[\, \tau-\frac{2}{3}\, \ln\frac{\tau}{4\pi}
+ B + \ln ck^2+o(1)\right]\,,
\label{MT3}
\end{equation}
where $B=1+\Delta_1+\Delta_3=-0.514362926\dots$,
$o(1)$ denotes terms that vanish as $\tau \to \infty$ and we have restored
the $k$-dependence \cite{k=1}. We see that the leading term in 
$a(\tau)$ is logarithmic in physical time $t$, and it coincides with
Eq. (\ref{leading}). The 
subleading term behaves like $\ln \tau \sim \ln \ln t$, while
the sub-subleading term is constant.

The matching region is determined by the requirements
that the subleading term in Eq. (\ref{QS14}) is much less than the 
leading term, but much greater than the rest of terms. These yield an
approximate condition $\ln^2 a \ll- x+\frac{\pi}{2}\,a\ll a^{2/3}$, so that
the matching
region expands as $\tau \to \infty$. 

\begin{figure}[h]
\vspace{-0.1in}
\hspace{0.0cm}
\rightline{ \epsfxsize = 8.6cm \epsffile{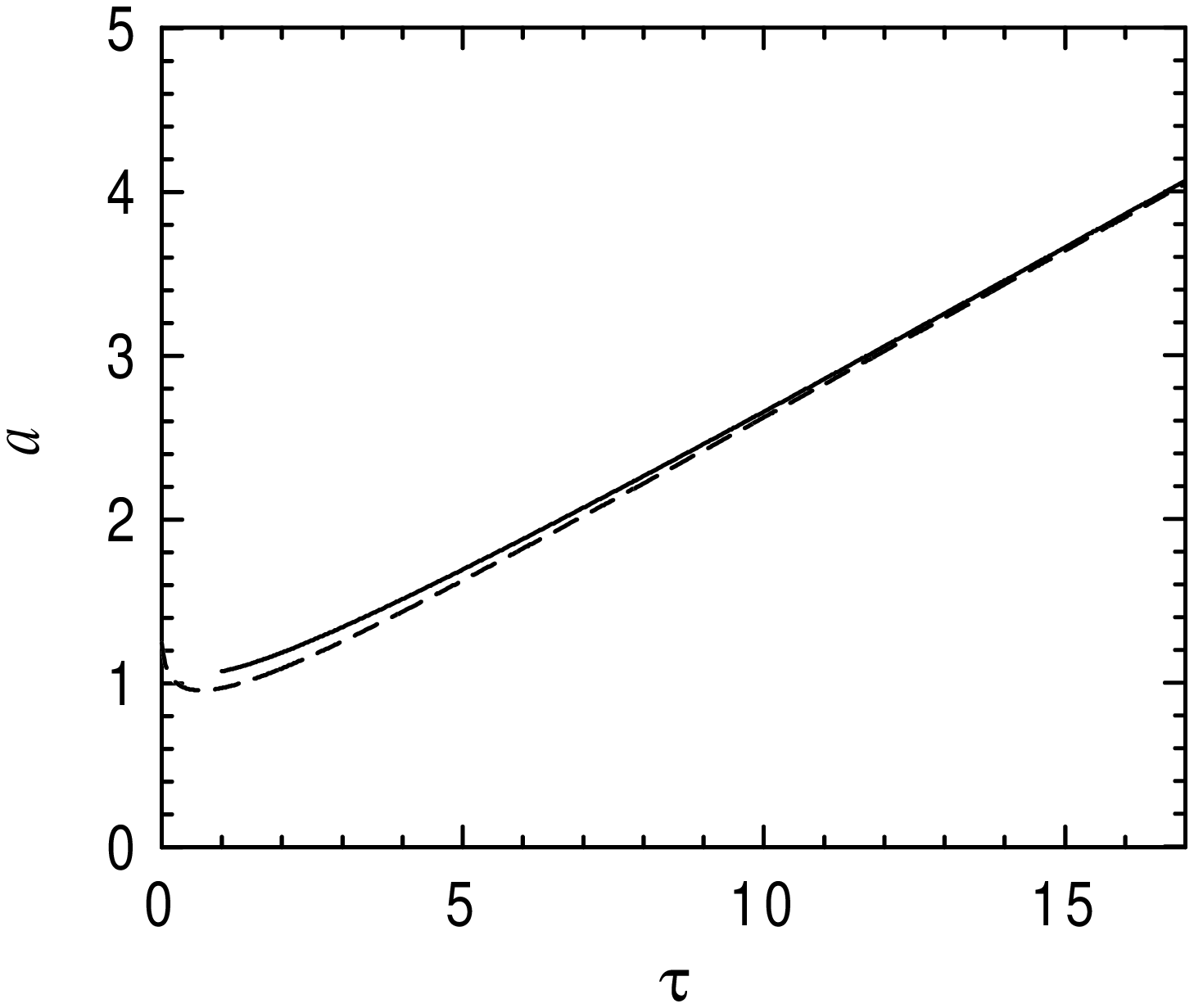}}
\vspace{0.1in}
\caption{The amplitude $a$ versus the new time $\tau= \ln t$
found numerically (dashed line) and analytically (solid line). The
parameters
are described in the text.
\label{fig. 1}}
\end{figure}

We compared the asymptotic solution with numerical simulations.
Finding 
logarithmic corrections numerically usually requires going to 
very long times. Instead, we directly solved Eq. (\ref{3}) in the 
new variables, which enabled us to 
reach $\tau \sim 20$, that is $t \sim 5 \cdot 10^8$. Eq.
(\ref{3}) was solved on the 
interval $x \in (0, 9)$ subject to the Neumann
boundary conditions. The initial condition was $u (x, \tau=0) =
2 \exp\left[-3\,(x^2+0.1)^{1/2}\right]$, so that $c=2$ and $k=3$.
The system length was 
large enough for the solution to enter
the asymptotic regime before the 
expanding ``core" reaches
the boundary $x=9$. Fig. 1 compares 
$a=\left(2 u(0,\tau)\right)^{1/2}$ found numerically 
with the
prediction of
Eq. (\ref{MT3}). At long times the agreement is excellent. We also verified
other attributes of the asymptotic solution.

In separate simulations that will be presented elsewhere \cite{KLM}, 
evolution of the same initial condition was investigated in the
framework of the {\it full gasdynamic} equations. Remarkably, the 
results essentially coincide, even for moderate Mach numbers, 
with those obtained 
with  Eq. (\ref{1}). This shows robustness of the reduced equation
in the description of CCFs. 

In conclusion, we have shown that 
hot bubbles in gases expand logarithmically slowly in the process
of cooling.  By constructing an asymptotic solution, 
that matches a ``quasi-similarity" inner solution
and a ``glassy" outer solution, we have been able to see how
logarithmic corrections enter dynamic scaling. 

We are grateful to Y. Kurzweil for help with Fig. 1. This work 
was partially supported
by the COE Visiting Research 
Scholar Program at YITP and by the
Russian Foundation for Basic Research (grant No. 99-01-00123).


\begin{references}
\bibitem[*]{adr} {On leave from the Institute of Theoretical and 
Experimental Physics, Moscow 117259, Russia.}
\bibitem{scaling}   A.J. Bray, 
Adv. Phys. {\bf 43}, 357 (1994); 
A.-L. Barab\'{a}si and H.E. Stanley, {\it Fractal Concepts
in Surface Growth}, (Cambridge Univ. Press, Cambridge, 1995);
N. Goldenfeld, {Lectures on Phase Transitions and
the Renormalization Group} (Addison-Wesley, 1992).
\bibitem{Barenblatt} G.I. Barenblatt, {\it Scaling, Self-similarity, and 
Intermediate Asymptotics} (Cambridge Univ. Press, Cambridge, 1996).
\bibitem{M} B. Meerson, Phys. Fluids A {\bf 1}, 887 (1989).
\bibitem{K} D. Kaganovich, B. Meerson, A. Zigler, C. Cohen and
J. Levin,  Phys. Plasmas {\bf 3}, 
631 (1996).
\bibitem{GLM} A. Glasner, E. Livne and B. Meerson, Phys. Rev. 
Lett. {\bf 78},
2112 (1997).
\bibitem{density} Alternatively, one obtains a nonlinear diffusion
equation for the gas density: $\partial_t \rho = 
\partial_x (\rho^{-\nu-1} \partial_x \rho)$, with the effective
diffusion coefficient
{\it decreasing} with $\rho$. This equation appeared in a number
of nonlinear diffusion problems where $\rho$ 
decays at $|x| \to \infty$ \cite{Rosenau}. In the hot 
bubble problem $\rho \to \infty$ at $|x| \to \infty$, and this difference
results in quite a different dynamics.
\bibitem{CH} M.C. Cross
and P.C. Hohenberg, Rev. Mod. Phys. {\bf 65}, 851 (1993).
\bibitem{ZR} Ya. B. Zel'dovich and Yu. P. Raizer, {\em The Physics
of Shock Waves and High Temperature Hydrodynamic Phenomena}
(Academic, New York, 1967).
\bibitem{shrinking} Shrinking is possible in the problem of 
{\it heating} of a cold ``cloud" by a hot and
underdense peripheral gas. Some aspects of this regime were
investigated in Ref. \onlinecite{Rosenau}.
\bibitem{intermediate} In reality, 
the temperature should approach
a {\it finite} value at large distances. Correspondingly,
the solution we are interested
in represents an {\it intermediate} asymptotics \cite{Barenblatt}.
\bibitem{k=1} If $u (x,\tau)$ is a solution of Eq. (\ref{3}), then 
$\lambda^2\,u (x/\lambda, \tau)$ is also a solution for any $\lambda>0$. 
This property enables one to restore the $k$-dependence in 
the final results. 
\bibitem{Mathematica} S. Wolfram, {\it The Mathematica Book}, 3rd edition
(Cambridge Univ. Press, Cambridge, 1996).
\bibitem{KLM} Y. Kurzweil, E. Livne and B. Meerson (unpublished).
\bibitem{Rosenau} P. Rosenau, Phys. Rev. Lett. {\bf 74}, 1056 (1995).
\end{references}
\end{document}